\documentclass[runningheads,a4paper]{llncs}
\usepackage{amssymb}
\usepackage{amsmath} 
\usepackage{graphicx}
\usepackage{subfigure}
\usepackage{multirow}
\usepackage{url}
\urldef{\mailsa}\path||
\urldef{\mailsb}\path||
\urldef{\mailsc}\path||    
\newcommand{\keywords}[1]{\par\addvspace\baselineskip
\noindent\keywordname\enspace\ignorespaces#1}

\begin{document}
\mainmatter  

% Title of the Paper
\title{Design of Novel Algorithm and Architecture for Gaussian Based Color Image Enhancement System for Real Time Applications}

% Auhtor Names and Affiliations
\author{M. C Hanumantharaju$^{1}$, M. Ravishankar$^{1}$ and  D. R Rameshbabu$^{2}$ \\
\vspace{5 mm} $^1$Department of Information Science \&\ Engineering, \\
              $^2$Department of Computer Science \&\ Engineering
              \vspace{2 mm} \\Dayananda Sagar College of Engineering, Bangalore, India.
\vspace{2 mm}\email{(mchanumantharaju, ravishankarmcn, bobrammysore)@gmail.com}}
\authorrunning{M. C Hanumantharaju et al.}

\maketitle

\begin{abstract}
This paper presents the development of a new algorithm for Gaussian based color image enhancement system. The algorithm has been designed into architecture suitable for FPGA/ASIC implementation. The color image enhancement is achieved by first convolving an original image with a Gaussian kernel since Gaussian distribution is a point spread function which smoothes the image. Further, logarithm-domain processing and gain/offset corrections are employed in order to enhance and translate pixels into the display range of 0 to 255. The proposed algorithm not only provides better dynamic range compression and color rendition effect but also achieves color constancy in an image. The design exploits high degrees of pipelining and parallel processing to achieve real time performance. The design has been realized by RTL compliant Verilog coding and fits into a single FPGA with a gate count utilization of 321,804. The proposed method is implemented using Xilinx Virtex-II Pro XC2VP40-7FF1148 FPGA device and is capable of processing high resolution color motion pictures of sizes of up to $1600\times1200$ pixels at the real time video rate of 116 frames per second. This shows that the proposed design would work for not only still images but also for high resolution video sequences.

\keywords{Gaussian color image enhancement, Serpentine memory, 2D Gaussian convolution, Logarithm, Field Programmable Gate Array.}
\end{abstract}

\section{Introduction}
Digital image enhancement \cite{Ref1} refers to accentuation, sharpening of image features such as edges, boundaries or contrast to make a graphic display more useful for display and analysis. The enhanced images with better contrast and details are required in many areas such as computer vision, remote sensing, dynamic scene analysis, autonomous navigation and medical image analysis. In the recent years, color image enhancement has been becoming an increasingly important research area with the widespread use of color images. Numerous methods \cite{Ref2} are available in the literature for color image enhancement. The color image enhancement can be classified into two categories according to the color space: color image enhancement in RGB color space and color image enhancement based on transformed space. 

The paper is organized as follows: Section 2 gives a brief review of previous work done. Section 3 describes the proposed Gaussian based color image enhancement algorithm. Section 4 provides detailed system architecture for hardware realization of color image enhancement algorithm. The results and discussions follow these. Conclusion arrived at are presented in the final section. 

\section{Related Work}
Digital Signal Processors (DSPs) \cite{Ref3}\cite{Ref4} have been employed for enhancement of images which provides some improvement compared to general purpose computers. Only marginal improvement has been achieved since parallelism and pipelining incorporated in the design are inadequate. This scheme uses optimized DSP libraries for complex operations and does not take full advantage of inherent parallelism of image enhancement algorithm. The neural network based learning algorithm \cite{Ref5} provides an excellent solution for the color image enhancement with color restoration. The hardware implementation of these algorithms parallelizes the computation and delivers real time throughput for color image enhancement. However, its window related operations such as convolution, summation and matrix dot products in an image enhancement architecture demands enormous amount of hardware resources. 

\par Hiroshi Tsutsui et al. \cite{Ref6} proposed an FPGA implementation of adaptive real-time video image enhancement based on variational model of the Retinex theory. The authors have claimed that the architectures developed in this scheme are efficient and can handle color picture of size $1900\times1200$ pixels at the real time video rate of 60 frames per sec. The authors have not justified how high throughput has been achieved in spite of time consuming iterations to the tune of 30. Abdullah M. Alsuwailem et al. \cite{Ref7} proposed a new approach for histogram equalization using FPGAs. Although efficient architectures were developed for histogram equalization, the restored images using this scheme are generally not satisfactory. 

\par An efficient architecture for enhancement of video stream captured in non-uniform lighting conditions was proposed by Ming Z. Zhang et al. \cite{Ref8}. The new architecture processes images and streaming video in the HSV domain with the homomorphic filter and converts the result back to HSV. This leads to an additional computational cost and, the error rate is high for the RGB to HSV conversion process. Digital architecture for real time video enhancement based on illumination reflection model was proposed by Hau T. Ngo et al. \cite{Ref9}. This scheme improves visual quality of digital images and video captured under insufficient and non-uniform lighting conditions. Bidarte et al. \cite{Ref10} proposed spatial-based adaptive and reusable hardware architecture for image enhancement. However, the histogram modification used in this scheme treats all regions of the image equally and often results in poor local performance, which in turn affects the image details. The modified luminance based multiscale retinex algorithm proposed by Tao et al. \cite{Ref11} achieves optimal enhancement result with minimal complexity of hardware implementation. However, the algorithm works fine so long as the background is dark and the object is bright. 

\par The limitations mentioned earlier are overcome in the proposed method in an efficient way. To start with, the input image is convolved with $5\times5$ Gaussian kernel in order to smooth the image. Further, the dynamic range of an image is compressed by replacing each pixel with its logarithm. In the proposed method, the image enhancement operations are arranged in an efficient way adding true color constancy at every step. It has less number of parameters to specify and provides true color fidelity. In addition, the proposed algorithm is computationally inexpensive. In the proposed scheme, an additional step is necessary to solve the gray world violation problem as is the case with the implementation reported in Ref. \cite{Ref11}. 

In the present work, in order to test the developed algorithm, standard test images have been used and results are favorably compared with that of other researchers. In order to evaluate the performance of the proposed algorithm, the metric Peak Signal to Noise Ratio (PSNR) has been used. 

\section{Proposed Gaussian Based Image Enhancement Algorithm}
In order to account for the smoothness, lightness, color constancy and dynamic range properties of Human Visual System (HVS), Gaussian based image enhancement algorithm has been proposed. The basic operational sequence of the proposed Gaussian based color image enhancement algorithm is shown in Fig. 1. To start with, the original image (which is of poor quality and needing enhancement) is read in RGB color space. The color components are separated followed by the selection of window size as $5\times5$ for each of the R, G and B components. In each color component, the selected window is convolved with $5\times5$ Gaussian kernel in order to smooth the image. Next, Logarithmic operation is accomplished in order to compress the dynamic range of the image. Finally, Gain/Offset adjustment is done in order to translate the pixels into the display range of 0 to 255. The separated R, G and B components are combined into composite RGB in order to obtain the enhanced image. 

\begin{figure}
\centering
\includegraphics[width=3.5 in]{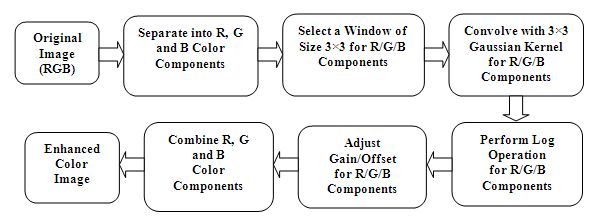}
\caption{Flow Sequence of the Proposed Gaussian based Color Image Enhancement Algorithm}
\label{fig_sim}
\end{figure}

\subsection{Convolution Operation}
Convolution is a simple mathematical operation which is essential in many image processing algorithms. The color image enhancement algorithm proposed in this paper uses $5\times5$ pixel window, where a block of twenty five pixels of original image is convolved with Gaussian kernel of size $5\times5$. The two dimension (2D) Gaussian function is defined by Eqn. (1): 

\begin{equation}
g(x, y) = \frac{1}{2\pi\sigma^{2}}e^{-\frac{x^{2} + y^{2}}{2\sigma^{2}}}
\end{equation}

where $\sigma$ is the standard deviation of the distribution; x and y are spatial coordinates.

The basic principle of Gaussian filters is to use 2D distribution as a point spread function. This property is adopted in image enhancement application and is achieved by convolving the 2D Gaussian distribution function with the original image. A Gaussian kernel can be a $3\times3$ or a $5\times5$ matrix as shown in Fig. 2. 
\begin{figure}
 \begin{center}
\begin{array}[t]{cc}
\centering
\frac{1}{16}\begin{array}{|l|c|r|} \hline
  1 & 2 & 1 \\ \hline
  2 & 4 & 2 \\ \hline 
  1 & 2 & 1 \\ \hline 
\end{array}			
& \hspace{1 cm}
\frac{1}{273}\begin{array}{|l|c|r|l|c|} \hline
  1 & 4  & 7  & 4  & 1     \\ \hline
  4 & 16 & 26 & 16 & 4     \\ \hline
  7 & 26 & 41 & 26 & 7     \\ \hline
  4 & 16 & 26 & 16 & 4     \\ \hline
  1 & 4  & 7  & 4  & 1     \\ \hline \end{array}
\end{array}
\caption{Gaussian Kernels: $3\times3$ kernel and $5\times5$ kernel}
\label{flowchart}
\end{center}
\end{figure}
%\vspace{0.1 cm}

The Gaussian convolution matrix is given by Eqn. (2).
\begin{equation}
G(x, y) = I(x, y) \otimes g(x, y)
\end{equation}

where $\otimes$ denotes convolution, g(x, y) is a Gaussian kernel, I(x, y) is the original image and G(x, y) is the convolved output. 

Mathematically, 2D convolution can be represented by the following Eqn. (3):

\begin{equation}
G(x, y) = \sum_{i=1}^{M}\sum_{j=1}^{N}{I(i, j)\times g(x-i, y-j)}
\end{equation}

where m = 5 and n = 5 for $5\times5$ Gaussian kernel. In this work, the convolution mask is chosen as $5\times5$ in order to enhance the implementation speed, at the same time minimizing blocking artifacts. The convolution operation for a mask of $5\times5$ is given by Eqn. (4)

\begin{equation}
P(x, y) = \frac{\sum_{i=0}^{4}W_i\times P_i}{\sum_{i=0}^{4}W_i}
\end{equation}

where $W_i$ indicates the $5\times5$ Gaussian mask, $P_i$ is the $5\times5$ sliding window in the input image and P is the Gaussian convolved pixel. As an example, Fig. 3 illustrated the hardware implementation of convolution with mask as $3\times3$. The convolution mask of $5\times5$ implemented in this work is similar to that shown in Fig. 3.

\begin{figure}
\centering
\includegraphics[width=3 in]{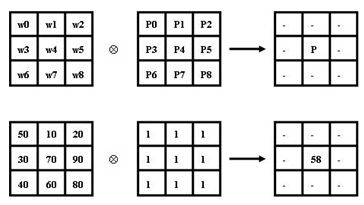}
\caption{Two Dimensional Convolution Scheme}
\label{fig_sim}
\end{figure}

Next, the log transformation operation applied on the image compresses the dynamic range of gray level input values to manageable levels for further processing. The logarithmic processing on a 2D image is carried out by using Eqn. (5):

\begin{equation}
G_L(x, y) = K\times\log_2\left[1 + G(x, y)\right]
\end{equation}

where K is a constant arrived as 1.5 after conducting a number of experiments with various test images. Further, the logarithm computed image is scaled by a scaling factor of 32 for getting pixel values in the range of 0 to 255. The final step in proposed approach is gain/offset correction. This gain/offset correction is accomplished by using Eqn. (6). 

\begin{equation}
I'(x, y) = \frac{d_{max}}{G_{Lmax}-G_{Lmin}}\left[G_L(x, y)-G_{Lmin}\right]
\end{equation}

where $d_{max}$ is the maximum intensity which is chosen as 255 for an image with 8-bit representation, $G_L(x,y)$ is the log-transformed image, $G_{Lmin}$  is the minimum value of log transformed image, $G_{Lmax}$  is the maximum value of log transformed image, $I'(x, y)$  is the enhanced image and, x and y represent spatial coordinates.

\section{Architecture of Proposed Color Image Enhancement System}
This section presents the architectures of the proposed color image enhancement system. It is composed of several components such as serpentine memory, 2D convolution, logarithm and gain/offset corrections. 
\subsection{System Overview}
The block diagram of a top level Gaussian based color image enhancement system shown in Fig. 4, which performs a color image enhancement operation as implied in the name. The detailed signal descriptions of this block are provided in Table I. The inputs "rin", "gin", and "bin" are the red, green, and blue pixels, each of size 8-bits. The input pixels are valid at the positive edge of the clock. The outputs "ro", "go" and "bo" represent the enhanced pixels of proposed image enhancement system. The "pixel\_valid" signal is asserted, When the enhanced pixel data is valid. The enhancement system can be reset at any point of time by asserting the asynchronous, active low signal, "reset\_n".

\begin{figure}
\centering
\includegraphics[width=2.5 in]{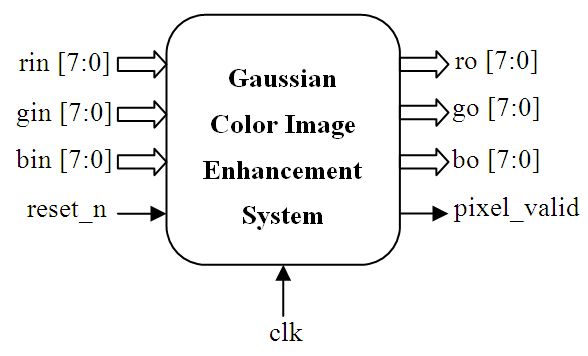}
\caption{Block Diagram of Top Level Gaussian Based Color Image Enhancement System}
\label{fig_sim}
\end{figure}

\begin{table}
\renewcommand{\arraystretch}{1.3}
\caption{Signal Description for the Top Module of Gaussian Based Image Enhancement System
}
\label{table_example}
\centering
% The array package and the MDW tools package offers better commands
% for making tables than plain LaTeX2e's tabular which is used here.
\begin{tabular}{|l|l|}
\hline
\bf Signals & \bf Description \\
\hline
clk & This is the global clock signal\\
\hline
reset\_n & Active low system reset  \\
\hline
rin [7:0] & Red color component\\
\hline
gin [7:0] & Green color component\\
\hline
bin [7:0] & Blue color component\\
\hline
ro [7:0] & Enhanced red color component\\
\hline
go [7:0] & Enhanced Green color component\\
\hline
bo [7:0] & Enhanced Blue color component\\
\hline
pixel\_valid & Valid signal for enhanced RGB pixel\\
\hline
\end{tabular}
\end{table}

The architecture proposed for color image enhancement system consists of twelve modules, each color component (R/G/B) comprising four basic modules: Sliding window, 2D Gaussian Convolution, Logarithm Base-2, and Gain/Offset Correction as shown in Fig. 5. Pipelining and parallel processing techniques have been adopted in the design in order to increase the processing speed of the proposed system. 

\begin{figure}
\centering
\includegraphics[width=4 in]{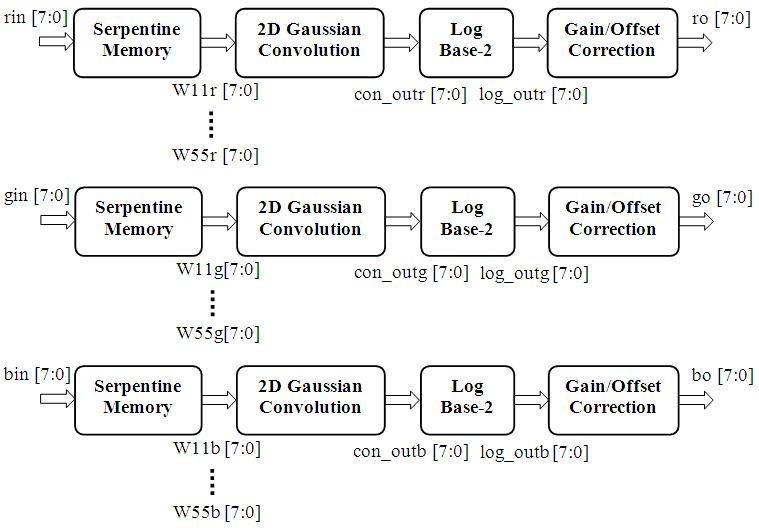}
\caption{Architecture of Gaussian Based Color Image Enhancement System}
\label{fig_sim}
\end{figure}

\subsection{Serpentine Memory Architecture}
The signal diagram of a serpentine memory for each of the color components R, G and B is presented in Fig. 6(a). The serpentine memory is commonly referred to as a sliding window. The proposed method uses $5\times5$ sliding window due to its ease of implementation. The pixel values of the input image "pixel\_in" of width 8-bits are imported serially into the sliding window module on the positive edge of clock, "clk". Detailed hardware architecture for $5\times5$ sliding window function or serpentine memory that uses row buffers is shown in Fig. 6(b).  

\begin{figure}
\centering
\subfigure[]{
\includegraphics[width=2 in]{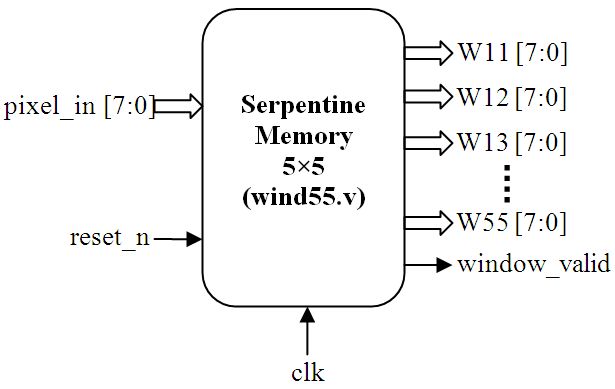}
\label{fig:subfig1}
}
\subfigure[]{
\includegraphics[width=2.5 in]{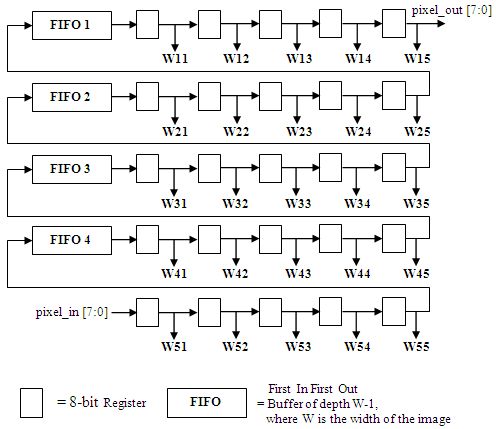}
\label{fig:subfig2}
}
\label{fig:subfigureExample}
\caption[Optional caption for list of figures]{(a) Signal Diagram of Serpentine Memory for R/G/B Color Channels (b) Schematic Diagram of a $5\times5$ Serpentine Memory for R/G/B Color Channels}
\end{figure}

%\begin{figure}
%\centering
%\includegraphics[width=3 in]{figures/fig6.jpg}
%\caption{Signal Diagram of Serpentine Memory for R/G/B Color Channels}
%\label{fig_sim}
%\end{figure}
%
%\begin{figure}
%\centering
%\includegraphics[width=4 in]{figures/fig7.jpg}
%\caption{Schematic Diagram of a $5\times5$ Serpentine Memory for R/G/B Color Channels}
%\label{fig_sim}
%\end{figure}

In this work, one pixel is read from memory in one clock cycle with no latency. The pixels are read row by row in a raster scan order. For a $5\times5$ sliding window, four First-In-First-Out (FIFO) buffers are used. The FIFO buffers are used to reduce the memory access to one pixel per clock cycle. The depth of the FIFO buffer is chosen as (W-3), where W is the width of the image. 

\subsection{Architecture of Gaussian Convolution Processor}
The design of the 2D Gaussian convolution processor in hardware is more difficult than that of the sliding window. The convolution algorithm uses adders, multipliers and dividers in order to calculate its output. The signal diagram of the 2D Gaussian convolution processor is shown in Fig. 7. 

\begin{figure}
\centering
\includegraphics[width=2.5 in]{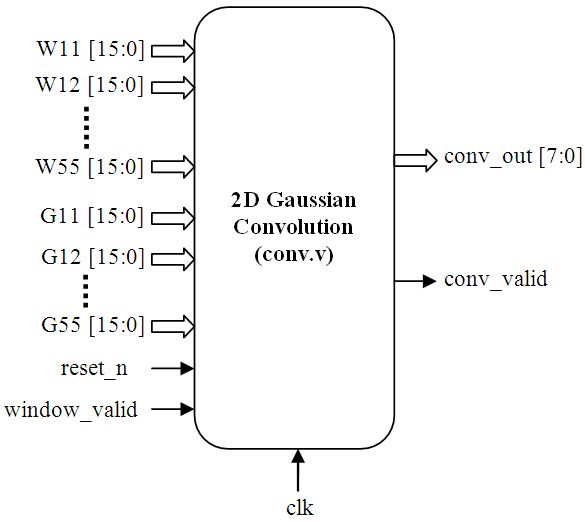}
\caption{Signal Diagram of 2D Gaussian Convolution Processor for R/G/B Color Channels}
\label{fig_sim}
\end{figure}

The input pixels W11 to W55 are scaled to 16 bits in order to match the Gaussian coefficients. The input pixels are valid on the positive edge of the "clk" with "window\_valid" asserted. The coefficient values of the convolution mask employed in most of the image processing applications remain constant for the entire processing. Therefore, constant coefficient multiplier is employed in order to implement convolution efficiently. The twenty five Gaussian coefficients G11 [15:0] to G55 [15:0] represent the convolution mask shown earlier in Fig. 2. The output signal "conv\_out" is 8-bits in width and is valid only on the positive edge of signal, "conv\_valid". 

\subsection{Logarithm of Base-2 Architecture}
The signal diagram of logarithm base-2 module for R/G/B color channels is shown in Fig. 8(a). The convolution output comprises the input for the logarithm block. The architecture for logarithmic module is shown in Fig. 8(b).

\begin{figure}
\centering
\subfigure[]{
\includegraphics[height=2.5cm, width=3cm]{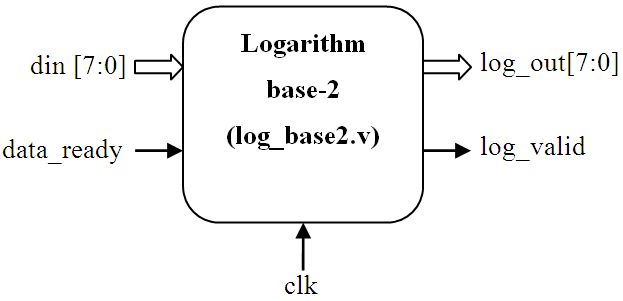}
\label{fig:subfig1}
}
\subfigure[]{
\includegraphics[height=3cm, width=4cm]{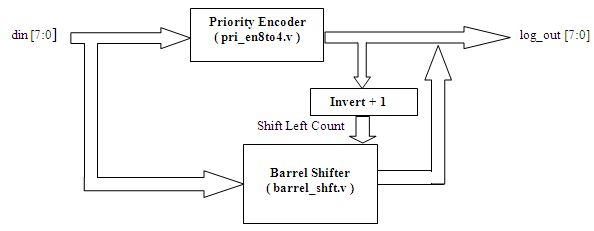}
\label{fig:subfig2}
}
\label{fig:subfigureExample}
\caption[Optional caption for list of figures]{(a) Signal Diagram of Logarithm Base-2 Module for R/G/B Color Channels (b) Architecture of Logarithm Base-2 Processor for R/G/B Color Channels}
\end{figure}

%\begin{figure}
%\centering
%\includegraphics[width=5 in]{figures/fig13.jpg}
%\caption{Architecture of Logarithm Base-2 Processor for R/G/B Color Channels}
%\label{fig_sim}
%\end{figure}

The log of base-2 is computed in two steps: Integer portion computation and fractional portion computation. Accordingly, the architecture consists of two modules. 

\subsection{Gain/Offset Correction}
The gain/offset correction is accomplished using Eqn. (6). The hardware implementation of gain and offset correction uses a multiplier and subtractor module. This step steers pixels into the correct display range: 0 to 255. The multiplier design presented in this work incorporates a high degree of parallel circuits and pipelining of five levels. The multiplier performs the multiplication of two 8-bits unsigned numbers n1 and n2 as shown in Fig. 9(a). The multiplier result is of width 16-bits. The detailed architecture for the multiplier is shown in Fig. 9(b). 

\begin{figure}
\centering
\subfigure[]{
\includegraphics[height=2.5cm, width=4cm]{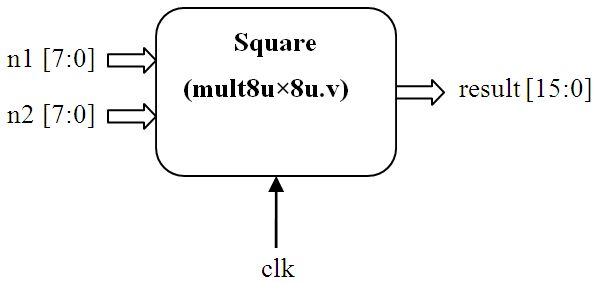}
\label{fig:subfig1}
}
\subfigure[]{
\includegraphics[height=4cm, width=5cm]{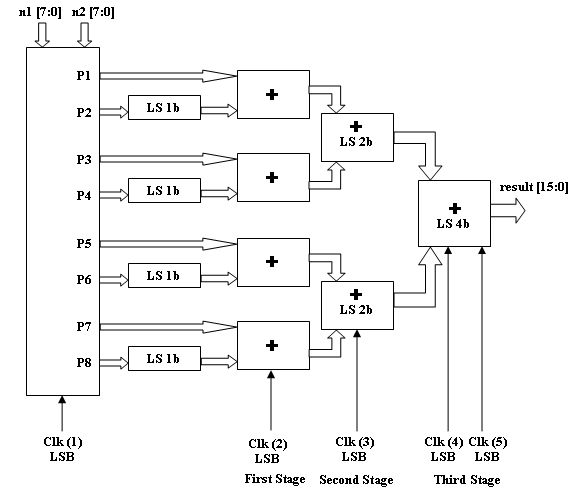}
\label{fig:subfig2}
}
\label{fig:subfigureExample}
\caption[Optional caption for list of figures]{(a) Signal Diagram of a Multiplier Block for R/G/B Color Channels(b) Detailed Architecture of Pipelined Multiplier Design for R/G/B Color Channels}
\end{figure}

%\begin{figure}
%\centering
%\includegraphics[width=5 in]{figures/fig15.jpg}
%\caption{Detailed Architecture of Pipelined Multiplier Design for R/G/B Color Channels}
%\label{fig_sim}
%\end{figure}

\section{Results and Discussions}
Over the years, many image enhancement techniques have been reported, major ones being Multiscale Retinex with Color Restoration. The Multiscale Retinex is the most popular technique that works well under most lighting conditions. In this paper, the most popular image enhancement techniques namely, histogram equalization\cite{Ref12}, MSRCR \cite{Ref13} and improved MSRCR \cite{Ref14} are chosen in order to validate the proposed scheme. 

The image enhancement algorithms mentioned earlier have been coded in Matlab by the present authors and the reconstructed images are shown in Fig. 10. The original image of resolution $256\times256$ pixels is shown in Fig. 10(a). The image enhancement using conventional histogram equalization of R, G and B channels is shown in Fig. 10(b). The histogram equalization method improves the global contrast of the original image. However, the overexposed regions of the original image are highly enhanced in this approach. Image enhancement based on Multiscale Retinex with Color Restoration is shown in Fig. 10(c). The enhanced image obtained by MSRCR is better compared with histogram equalization. MSRCR method fails to produce good color rendition for a class of images that contain violation of gray world assumption. 

The image enhanced based on improved multiscale retinex using hue-saturation-value (HSV) color space shown in Fig. 10(d) improves the visual quality better than MSRCR method. However, this approach is complex from computation point of view. The limitations expressed in the above algorithms are conspicuous by its absence in the proposed Gaussian based color image enhancement method as can be seen from the result displayed in Fig. 10(e). The proposed algorithm outperforms the other image enhancement techniques in terms of quality of the reconstructed picture. The images enhanced by our method are clearer, more vivid, and more brilliant than that achieved by other researchers.

\begin{figure}
\centering
\subfigure[]{
\includegraphics[scale=0.3]{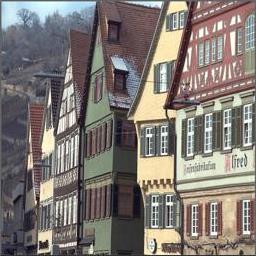}
\label{fig:subfig1}
}
\subfigure[]{
\includegraphics[scale=0.3]{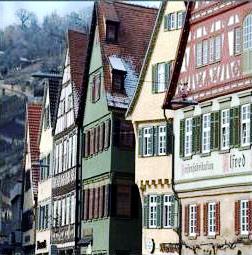}
\label{fig:subfig2}
}
\subfigure[]{
\includegraphics[scale=0.3]{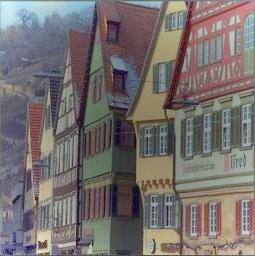}
\label{fig:subfig3}
}
\subfigure[]{
\includegraphics[scale=0.3]{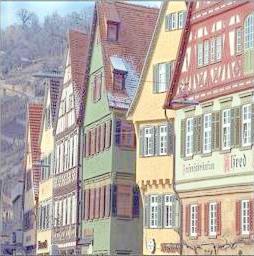}
\label{fig:subfig3}
}
\subfigure[]{
\includegraphics[scale=0.3]{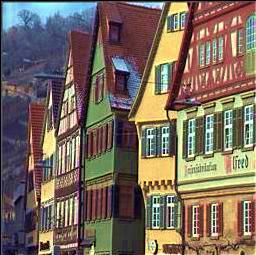}
\label{fig:subfig3}
}
\label{fig:subfigureExample}
\caption[Optional caption for list of figures]{Comparison of Reconstructed Pictures Using Image Enhancement Algorithms: (a) Original Image of resolution $256\times256$ pixels (b) Image enhanced based on Histogram Equalization of Ref. \cite{Ref12}, PSNR: 31.2 dB (c) Multiscale Retinex with Color Restoration of Ref. \cite{Ref13}, PSNR: 30 dB (d) Improved Multiscale Retinex in HSV Color Space of Ref. \cite{Ref14}, PSNR: 29.5 dB (e) Proposed Gaussian Color Image Enhancement, PSNR: 31.8 dB.}
\end{figure}

The proposed FPGA implementation of Gaussian based Color Image Enhancement System has been coded and tested in Matlab (Version 8.1) first in order to ensure the correct working of the algorithm. Subsequently, the complete system has been coded in Verilog HDL so that it may be implemented on an FPGA or ASIC. The proposed scheme has been coded in RTL compliant Verilog and the hardware simulation results are compared with Matlab results described earlier in order to validate the hardware design. The system simulation has been done using ModelSim (Version SE 6.4) and Synthesized using Xilinx ISE 9.2i. The algorithm has been implemented on Xilinx Virtex-II Pro XC2VP40-7FF1148 FPGA device. In the proposed work, window size is chosen as $5\times5$ since the enhanced image looks more appealing than that for other sizes. 

Elaborate experiments were conducted on over a dozen varieties of images and consistently good results have been obtained for the proposed Verilog implementation. As examples, three poor quality images have been enhanced using the Gaussian based color image enhancement hardware system and is presented in Fig. 11. Test images (a), (d) and (g) of Fig. 11 show the original image of resolution $256\times256$ pixels. The images (b), (e) and (h) of Fig. 13 show the enhanced images using the proposed method based on Matlab approach. The images in Fig. 11 (c), (f) and (i) show the enhanced images using the proposed Gaussian method based on hardware (Verilog) approach.   

\begin{figure}
\centering
\subfigure[]{
\includegraphics[scale=0.3]{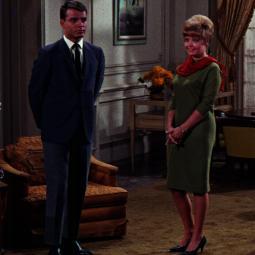}
\label{fig:subfig1}
}
\subfigure[]{
\includegraphics[scale=0.3]{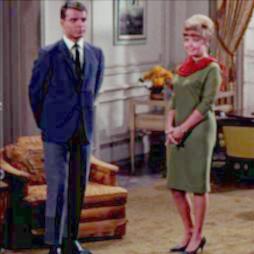}
\label{fig:subfig2}
}
\subfigure[]{
\includegraphics[scale=0.3]{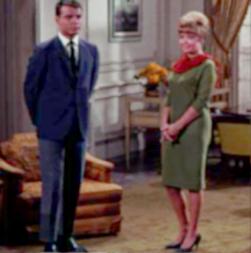}
\label{fig:subfig3}
}
\subfigure[]{
\includegraphics[scale=0.3]{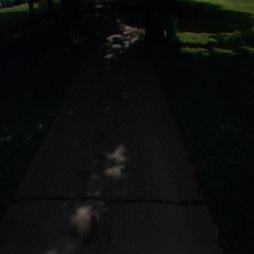}
\label{fig:subfig3}
}
\subfigure[]{
\includegraphics[scale=0.3]{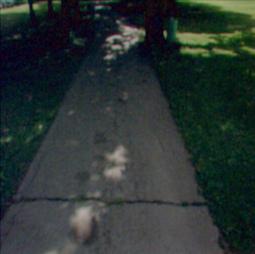}
\label{fig:subfig3}
}
\subfigure[]{
\includegraphics[scale=0.3]{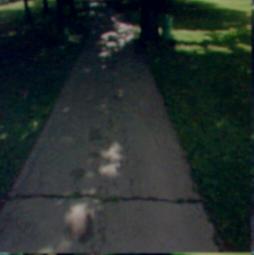}
\label{fig:subfig3}
}
\subfigure[]{
\includegraphics[scale=0.3]{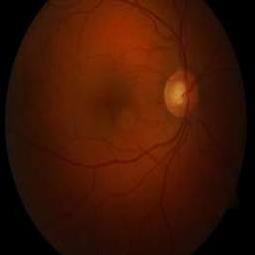}
\label{fig:subfig3}
}
\subfigure[]{
\includegraphics[scale=0.3]{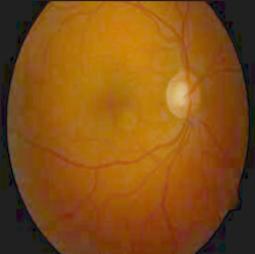}
\label{fig:subfig3}
}
\subfigure[]{
\includegraphics[scale=0.3]{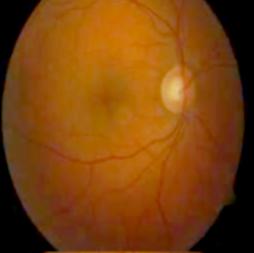}
\label{fig:subfig3}
}
\label{fig:subfigureExample}
\caption[Optional caption for list of figures]{Experimental Results for Gaussian based Color Image Enhancement System: (a) Original couple image ($256\times256$ pixels) (b) Reconstructed couple image using Matlab, PSNR: 36.9 dB (c) Reconstructed couple image using Verilog, PSNR: 36.2 dB. (d) Original dark road image ($256\times256$ pixels) (e) Reconstructed dark road image using Matlab, PSNR: 40.8 dB (f) Reconstructed dark road image using Verilog, PSNR: 40.3 dB. (g) Original eyeball image ($256\times256$ pixels) (h) Reconstructed eyeball image using Matlab, PSNR: 36.8 dB (i) Reconstructed eyeball image using Verilog, PSNR: 36.1 dB}
\end{figure}

\subsection{Verilog Simulation Results Using Modelsim}

The ModelSim simulation waveforms for inputting the image pixel data for all the three color components is shown in Fig. 12(a). The enhancement process starts when the "din\_valid" signal is asserted at the positive edge of the clock. The RGB image is separated into R, G and B data and the algorithm is applied concurrently to all the color components. The reconstructed pixels are issued at the output pins "dout" with "dvm" signal asserted as presented in Fig. 12(b). The outputs are issued out continuously one pixel every clock cycle after a latency of 535 clock cycles.

\begin{figure}
\centering
\subfigure[]{
\includegraphics[height=3.3cm, width=4.2cm]{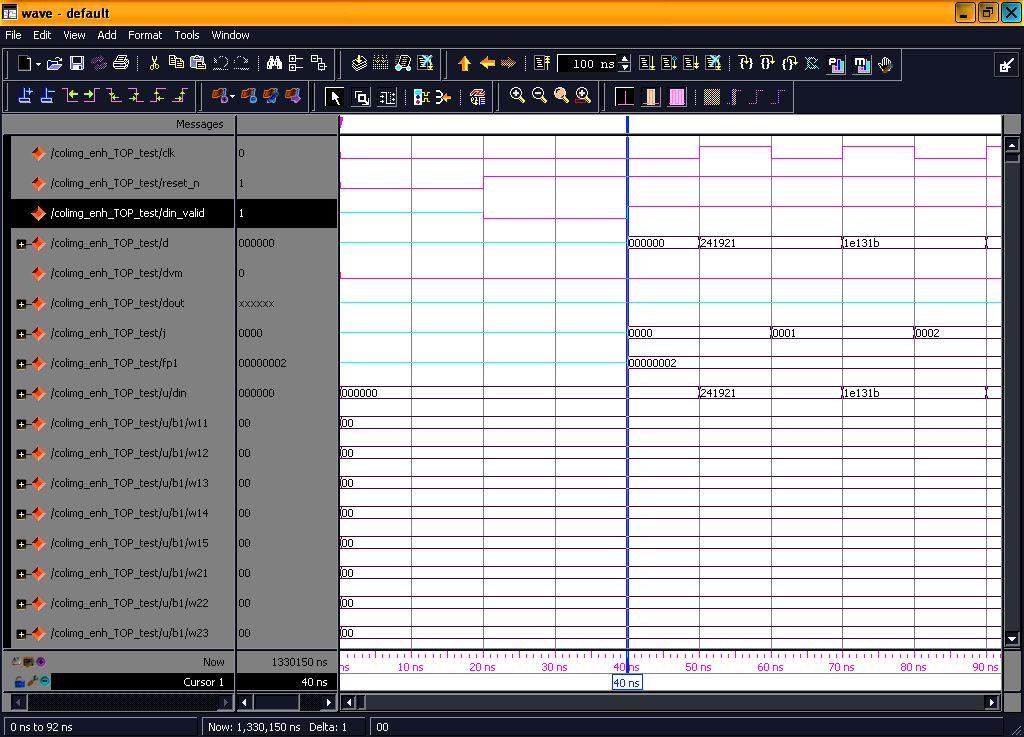}
\label{fig:subfig1}
}
\subfigure[]{
\includegraphics[height=3.3cm, width=4.2cm]{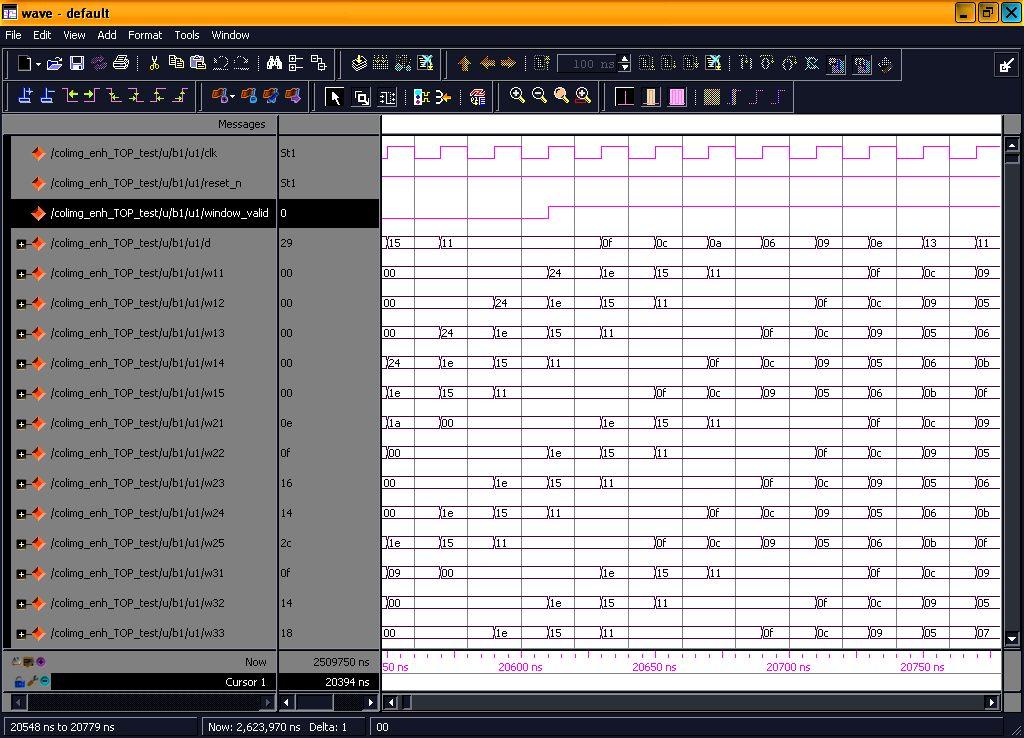}
\label{fig:subfig2}
}
\label{fig:subfigureExample}
\caption[Optional caption for list of figures]{(a) Waveforms for Gaussian Based Color Image Enhancement System: Inputting Image Data (b) Waveforms for Gaussian based Color Image Enhancement System: Enhanced Pixel Data}
\end{figure}

%
%\begin{figure}
%\centering
%\includegraphics[width=3 in]{figures/fig18.jpg}
%\caption{Waveforms for Gaussian Based Color Image Enhancement System: Inputting Image Data}
%\label{fig_sim}
%\end{figure}
%
%\begin{figure}
%\centering
%\includegraphics[width=5 in]{figures/fig19.jpg}
%\caption{Waveforms for Gaussian based Color Image Enhancement System: 
%Sliding Window Data for one of the Color Components
%}
%\label{fig_sim}
%\end{figure}

%\begin{figure}
%\centering
%\includegraphics[width=3 in]{figures/fig24.jpg}
%\caption{Waveforms for Gaussian based Color Image Enhancement System: Enhanced Pixel Data
%}
%\label{fig_sim}
%\end{figure}

%\begin{figure}
%\centering
%\includegraphics[width=5 in]{figures/fig23.jpg}
%\caption{Waveforms for Gaussian based Color Image Enhancement System:Multiplication Operation Results
%}
%\label{fig_sim}
%\end{figure}

%\begin{figure}
%\centering
%\includegraphics[width=3 in]{figures/fig24.jpg}
%\caption{Waveforms for Gaussian based Color Image Enhancement System: Enhanced Pixel Data
%}
%\label{fig_sim}
%\end{figure}

\subsection{Place and Route Results}
The design was Synthesized, Placed and Routed using Xilinx ISE 9.2i. The FPGA target device chosen was Xilinx Virtex-II Pro XC2VP40-7FF1148. The synthesis and place and route results for the design are presented in the following. The ISE generated RTL view of the top level module "Gaussian\_IE" is shown in Fig. 13(a) and the zoomed top view modules are presented in Fig. 13(b). The detailed zoomed view of R/G/B processor is shown in Fig. 14.

\begin{figure}
\centering
\subfigure[]{
\includegraphics[height=2.5cm, width=3cm]{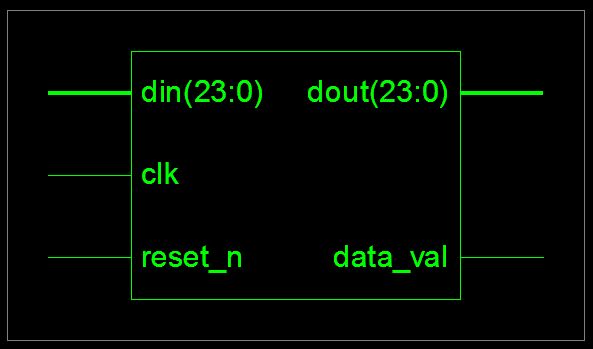}
\label{fig:subfig1}
}
\subfigure[]{
\includegraphics[height=3cm, width=5.5cm]{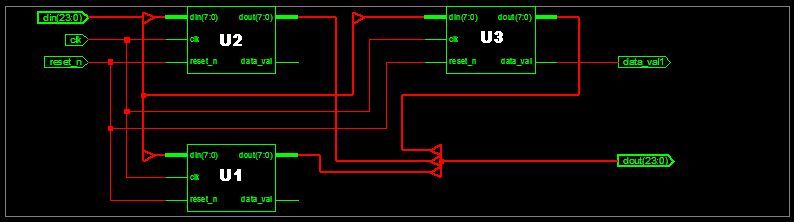}
\label{fig:subfig2}
}
\label{fig:subfigureExample}
\caption[Optional caption for list of figures]{Experimental Results for Gaussian based Color Image Enhancement System: (a) RTL View of the Top Module "Gaussian\_IE" (b) Zoomed View of "Gaussian\_IE" Module where \textbf{U1:} Red Color, \textbf{U2:} Green Color and \textbf{U3:} Blue Color Component Processors}
\end{figure}

\begin{figure}
\centering
\includegraphics[width=2 in]{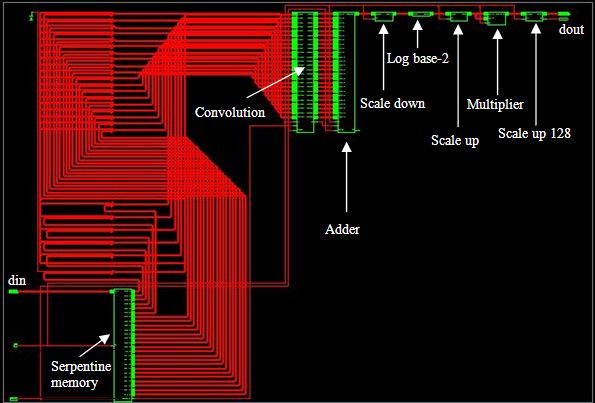}
\caption{Zoomed View of U1 or U2 or U3 Module.}
\label{fig_sim}
\end{figure}
%
%\begin{figure}
%\centering
%\includegraphics[width=5 in]{figures/fig26.jpg}
%\caption{Zoomed View of "Gaussian\_IE" Module where \textbf{U1:} Red Color, \textbf{U2:} Green Color and \textbf{U3:} Blue Color Component Processors}
%\label{fig_sim}
%\end{figure}
%
%\begin{figure}
%\centering
%\includegraphics[width=4 in]{figures/fig27.jpg}
%\caption{Zoomed View of U1 or U2 or U3 Module}
%\label{fig_sim}
%\end{figure}

The device utilization and the performance summary for Gaussian based color image enhancement system are presented in Fig. 15. The system utilizes about 321,804 gates as reported by the ISE tool. The maximum frequency of operation is 224.840 MHz. This works out to a frame rate of 117 per second for a picture size of $1600\times1200$ pixels since the design has the capability of processing one pixel every clock cycle ignoring initial latency of 535 clock cycles. As a result, it can work on any FPGA or ASIC without needing to change any code. As ASIC, it is likely to work for higher resolutions beyond 4K format at real time rates. This shows that the design would work for not only still images but also for high resolution video sequences.

\begin{figure}
\centering
\includegraphics[height=3cm, width=6cm]{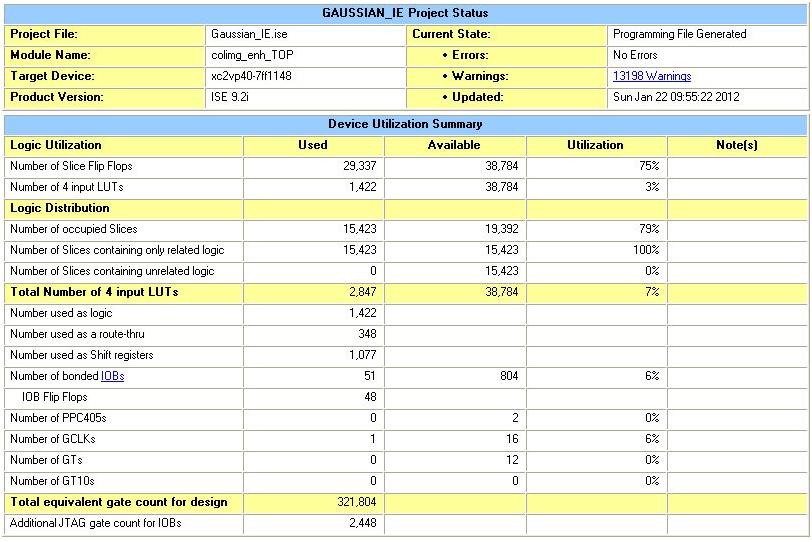}
\caption{FPGA Resource Utilization for Gaussian Based Color Image Enhancement Design
}
\label{fig_sim}
\end{figure}

The timing summary for the design as reported by Xilinx ISE tool is as follows:
\\Speed Grade : -7
\\Minimum period: 4.448ns (Maximum Frequency: 224.840 MHz)
\\Minimum input arrival time before clock:     1.418ns
\\Maximum output required time after clock:  3.293ns
\\Clock period: 4.448ns (frequency: 224.840 MHz)
\\Total number of paths /destination ports: 68988 / 30564
\\Delay: 4.448ns (Levels of Logic = 12)

\section{Conclusion}
A new algorithm and architecture for Gaussian based color image enhancement system for real time applications has been presented in this paper. The Gaussian convolution used in this scheme not only smoothes the image but also removes the noise present in the image. Further, the pixels are processed in log-domain in order compress the dynamic range. Finally, linear operation such as gain/offset correction is used in order to obtain image in display range. The Verilog design of the color image enhancement system is implemented on Xilinx Virtex-II Pro XC2VP40-7FF1148 FPGA device and is capable of processing high resolution videos up to $1600\times1200$ pixels at 117 frames per second.  The implementation was tested with various images and found to produce high quality enhanced images. The design realized using RTL compliant Verilog fits into a single FPGA chip with a gate count utilization of about 321,804.

\end{document}